\documentclass[sn-mathphys]{sn-jnl}
\usepackage{latexsym}
\usepackage{graphicx}
\usepackage{subfigure}
\usepackage{amsfonts,amssymb,amsmath}
\usepackage{hyperref}
\usepackage{array}
\usepackage{booktabs}
\usepackage{multirow}
\usepackage{makecell}
\usepackage{enumerate}

\jyear{2022}%

\theoremstyle{thmstyleone}%
\theoremstyle{thmstyletwo}%

\theoremstyle{thmstylethree}%

\begin{document}

\title[]{A Survey on Collaborative DNN Inference for Edge Intelligence}

\author[1]{\fnm{Wei-Qing} \sur{Ren}}

\author[1]{\fnm{Yu-Ben} \sur{Qu}}

\author[1]{\fnm{Chao} \sur{Dong}}

\author[1]{\fnm{Yu-Qian} \sur{Jing}}

\author[1]{\fnm{Hao} \sur{Sun}}

\author[1]{\fnm{Qi-Hui} \sur{Wu}}

\author[2]{\fnm{Song} \sur{Guo}}

\affil[1]{\orgdiv{College of Electronic and Information Engineering}, \orgname{Nanjing University of Aeronautics and Astronautics}, \orgaddress{\city{Nanjing},  \country{China}}}

\affil[2]{\orgdiv{Department of Computing}, \orgname{The Hong Kong Polytechnic University}, \orgaddress{\city{Hong Kong},  \country{China}}}

\abstract{With the vigorous development of artificial intelligence (AI), the intelligent applications based on deep neural network (DNN) change people's lifestyles and the production efficiency. However, the huge amount of computation and data generated from the network edge becomes the major bottleneck, and traditional cloud-based computing mode has been unable to meet the requirements of real-time processing tasks. To solve the above problems, by embedding AI model training and inference capabilities into the network edge, edge intelligence (EI) becomes a cutting-edge direction in the field of AI. Furthermore, collaborative DNN inference among the cloud, edge, and end device provides a promising way to boost the EI. Nevertheless, at present, EI oriented collaborative DNN inference is still in its early stage, lacking a systematic classification and discussion of existing research efforts. Thus motivated, we have made a comprehensive investigation on the recent studies about EI oriented collaborative DNN inference. In this paper, we firstly review the background and motivation of EI. Then, we classify four typical collaborative DNN inference paradigms for EI, and analyze the characteristics and key technologies of them. Finally, we summarize the current challenges of collaborative DNN inference, discuss the future development trend and provide the future research direction.}


\maketitle

\section{Introduction}\label{sec1}
In recent years, as the core technology of modern artiﬁcial intelligence (AI) and machine learning (ML) \cite{lecun2015deep}, deep neural network (DNN) represents the most commonly used ML technology and has a board range of intelligent applications such as intelligent assistant \cite{belkadi2020intelligent}, smart city \cite{bhattacharya2022review}, auto pilot \cite{rzadca2020autopilot}, intelligent agriculture \cite{ashifuddinmondal2018iot} and smart home \cite{pal2018internet}. While DNN has a great advantage in processing computation-intensive tasks, it relies heavily on sensors and other end devices to collect application data. With the rapid development of Internet of Things (IoTs) technology, the number of connected devices over the world becomes more and more huge, and each device hosts a large number of applications, which leads to the explosive growth of data at the edge \cite{mao2017survey}. Under this trend, the traditional cloud-based intelligence by first transferring data or tasks to the cloud center and then running DNN-based inference puts a heavy burden on the transmission links, which also results in unacceptable large responsive latency and growing data privacy disclosure concerns \cite{pu2015low}.

Bearing in mind those shortcomings of cloud-based intelligence, it is envisioned that the intelligence should be pushed from the cloud to the network edge, which gives birth to edge intelligence (EI) \cite{zhou2019edge}. In the service mode of EI, AI combines with edge computing, thus the network core sinks from the cloud to the edge closer to the data sources. It can make full use of edge resources to realize the wide DNN applications of AI. Meanwhile, edge nodes can connect to nearby terminal devices, servers, and gateways, and even micro base stations that can be used by nearby devices through device to device communication. Compared with the traditional cloud-based intelligence mode, employing EI to conduct DNN applications brings numerous advantages, including low response latency, high energy efficiency, privacy protection, bandwidth consumption reduction, throughput improvement, on-demand deployment, and well adaptation to some extreme scenario applications \cite{shi2016edge}.

EI could be generally categorized into two forms: edge training by embedding the training abilities over the edge nodes, and edge inference by deploying ML model inference on the edge nodes, where the latter is the main focus of this paper. By directly executing ML models such as DNN at the edge, edge inference could support relatively high-reliability and low-latency AI services through requiring less communication, computation, and storage resources. In general, edge inference could be divided into single node inference and multi-node collaborative inference. As ML models are developing towards deeper neural networks with higher computational requirements \cite{xu2018scaling}, it is challenging to implement complex models on a single resource-constrained edge device. Therefore, for large-scale inference tasks, EI oriented collaborative DNN inference \cite{letaief2021edge} can divide the original complex large-scale DNN inference task into different subtasks, and then dynamically allocate those subtasks to different nodes according to the computing power and energy of edge devices. In a nutshell, compared to single-node inference, multi-node collaborative inference can bring many potential benefits to the landing of EI, \textit{e.g.}, lower latency and bandwidth pressure, broader application scenarios, lower energy consumption and equipment rental cost, and agile task scheduling strategies and resource allocation among the involved nodes \cite{park2019wireless,jang2019introduction}.

While multi-node collaborative inference is a promising way to realize EI, it is inevitably faced with several challenges as follows. Firstly, how to rationally partition the DNN model into several submodels to suit the heterogeneity of involved nodes ranging from the powerful cloud to resource-constrained IoT devices. Secondly, different AI services have diverse application requirements (\emph{e.g.}, minimizing inference latency and/or energy consumption, and maximizing inference accuracy), then how to schedule the collaboration together with the network resources is extremely challenging. Although there exist some studies about collaborative DNN inference for EI recently, there is still a lack of a special summary and discussion on the collaborative DNN inference paradigms with different composition and future research trends. Therefore, in order to further promote the development of EI, this paper makes a comprehensive investigation on the research results of collaborative DNN inference in recent years. Specifically, we firstly review the background of EI. Then we discuss the motivation, definition, and classification of collaborative inference. Next, we further review and classify the collaborative inference paradigms with different structures, as well as their optimization objectives and efficient technologies. Finally, we elaborate several challenges and opportunities in collaborative DNN inference for EI.

The rest of this paper is organized as follows.

\begin{enumerate}[1)]
  \item Section \ref{sec2} summarizes the basic concepts of EI and collaborative DNN inference, and divides the collaborative DNN inference into four categories according to different organizational structures.
  \item From Section \ref{sec3} to Section \ref{sec6}, we respectively summarize the architecture, key technologies, and optimization objectives of the four collaborative DNN inference paradigms in detail.
  \item Section \ref{sec7} discusses the existing challenges and puts forward several future research directions of collaborative DNN inference.
  \item Section \ref{sec8} concludes this paper.
\end{enumerate}

\section{Preliminaries and classification of collaborative DNN inference}\label{sec2}
In this section, we firstly introduce the definition of EI and its application advantages, and compare the EI boosted by collaborative DNN inference with the intelligence by traditional cloud-based method. Then we analyze the reasons why scholars put forward collaborative DNN inference for EI as a problem solution. Finally, we summarize and classify the concepts and working methods of four collaborative DNN inference paradigms.

\subsection{Introduction to EI}
EI is a new concept born from the combination of AI and edge computing \cite{zhou2019edge}. Its broad definition is a method of building intelligent machines to perform tasks like humans. AI is a theory, technology and application system for simulating and extending human intelligence. At present, a more general definition of edge computing is that edge computing refers to a new computational model that provides calculation services at the edge of the network \cite{shi2016edge}. The edge refers to any resource from the data source to the cloud computing center, including network resources. The edge computing is a continuum, and anything other than a cloud can be called an edge. The basic principle of edge computing is to migrate computing tasks to the edge devices that generate raw data. From the perspective of scope, edge computing is similar to fog computing. However, fog computing \cite{bonomi2012fog} has a hierarchical and network architecture, while edge computing focuses on individual nodes that do not form a network.

The combination of edge computing and AI produces EI, which is natural and inevitable, because there is a cross interactive relationship between them. On one hand, AI provides technologies and methods for edge computing, on the other hand, edge computing provides scenes and platforms for AI \cite{deng2020edge}. Edge computing aims to coordinate multiple collaborative edge devices and servers to process the generated data near edge devices. While AI aims to simulate intelligent human behavior by learning from data. Pushing AI to the edge could not only get the advantages of edge computing, such as lower inference latency and less communication consumption, but also improve the overall performance of AI and IoT systems.

EI can assist in processing the data generated at the edge of the network and release its operation potential. Due to the surge in the number and types of mobile devices and IoT devices, continuous perception of the real physical environment on the device side will produce a large amount of data, such as audio and video data. AI is necessary in this case, because it can quickly analyze huge amounts of data and extract feature information from them, so as to make high-quality decision-making behavior in the future. Deep learning is one of the most popular AI technologies which brings the ability of automatical patterns recognition and detecting abnormal data from edge devices. Then the effective information extracted from the sensing data is fed back to the server for real-time prediction and decision-making, such as public transport planning \cite{chen2015deepdriving}, intelligent city monitoring \cite{jouhari2021distributed} and forest fire early warning \cite{kalatzis2018edge}, so as to respond to the rapidly changing environment and improve the efficiency of scene application.

Edge computing enhances AI through data and application scenarios, so as to promote the development of EI. It is generally believed that the rapid development of deep learning is supported by four aspects \cite{zhou2019edge}: hardware, algorithm, data, and application scenarios. The improvement of hardware affects the operation ability of the whole system from the underlying architecture. The algorithm affects the efficiency of deep learning from the top. However, the role of data and application scenarios is not intuitive enough, causing them to be ignored. Generally speaking, we can improve the performance of deep learning algorithm by adding more layers of neurons to DNN, which will increase the amount of parameters and data of DNN. It can be seen that data also has a very important impact on the development of AI. The traditional data processing method is to process and store data in a large-scale cloud data center. However, with the rapid development of IoT, the disadvantages of traditional methods are becoming visible. If a large amount of data generated at the edge is handed over to the AI algorithms of the cloud data processing center, it will consume large bandwidth resources. In order to solve these problems, EI improves the overall performance of the system and reduces the latency of data processing by dispersing the computing power from the cloud center to the edge.

\begin{figure}[h]
\centerline{\includegraphics[width=\textwidth]{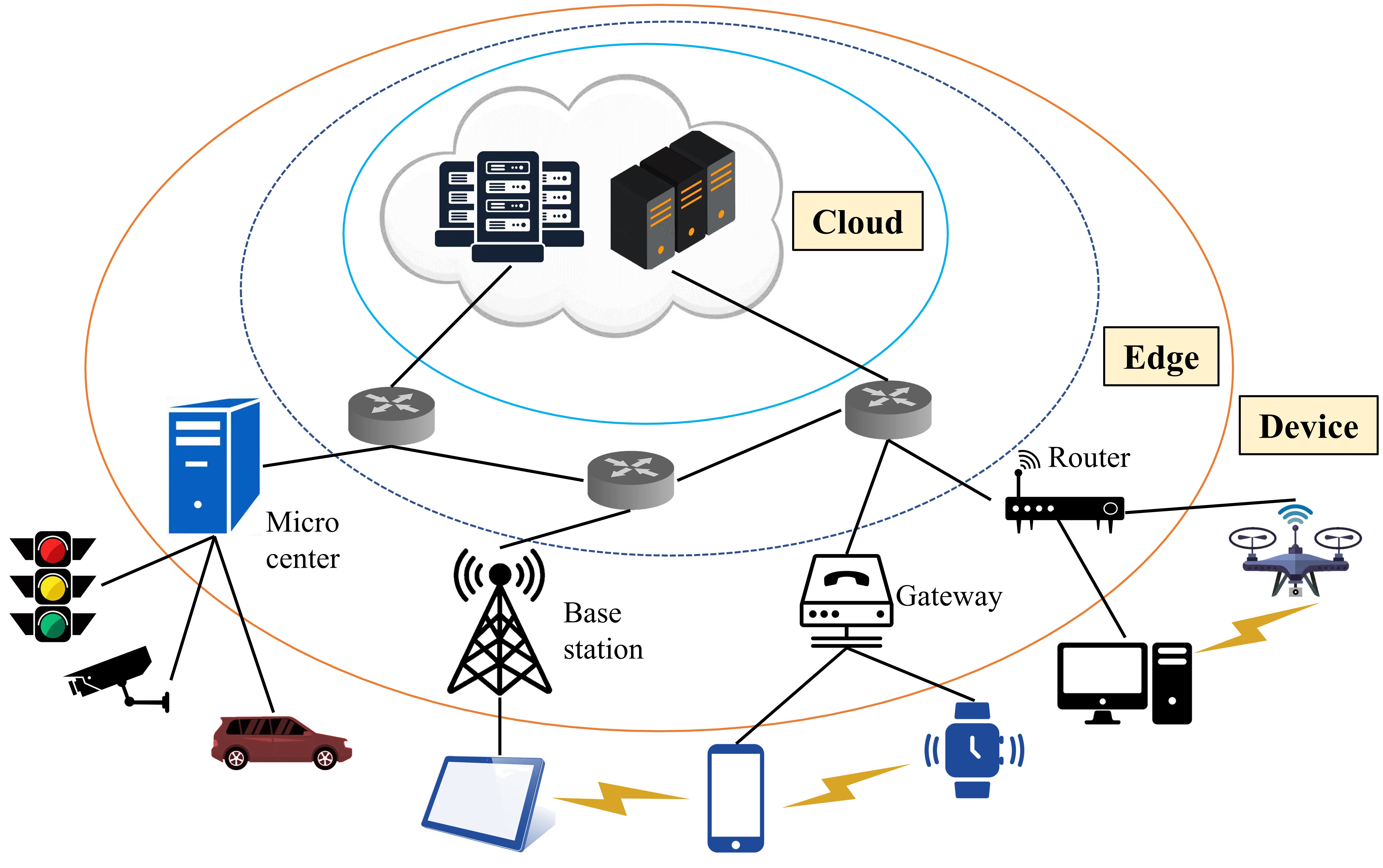}}
\caption{An illustration of the existing cloud-edge-device network structure \cite{zhou2019edge}.\label{fig1}}
\end{figure}

EI system has abundant edge resources. As shown in Fig.~\ref{fig1}, the cloud center connects edge servers through the WAN, such as base stations, gateways, routers, and micro centers. Then, these edge servers connect a large number of mobile terminal devices thereby build the cloud-edge-device network. And a large number of devices can be used to deploy DNN models to achieve collaborative inference. Meanwhile, the combination of devices at different levels will constitute different collaborative inference paradigms with different emphasis directions, which can be used in different application scenarios.

\subsection{Collaborative DNN inference for EI}
A DNN model consists of many different layers, such as convolution layer and full connection layer. Each layer converts its input according to the model parameters of DNN, and then passes through a nonlinear activation function, such as ReLU. As these layers process input and extract feature information in turn, they gradually establish high-level semantic information until the final prediction is generated.

With the continuous increase of mobile network bandwidth, multimedia interactive applications on mobile devices are growing rapidly, which involves intensive target recognition \cite{ren2015faster} \cite{liu2016ssd} \cite{redmon2017yolo9000} and image classification tasks \cite{szegedy2015going} \cite{he2016deep}. Convolutional neural network (CNN), as the representative of image information processing in deep neural network, is widely used in these tasks because of its high precision and high efficiency. As shown in Table.~\ref{tab1}, this paper sorts out some popular DNN models, including model type, parameter quantity, memory occupation and Giga floating point operations per second (GFLOPs). As the representative of the most classic CNN structure, VGG \cite{simonyan2014very} network won the second place in the 2014 ILSVRC competition. However, executing the DNN model requires a lot of computing and memory resources. VGG-16 has 140 million parameters, occupies more than 500 MB of memory, and requires 15.5 GFLOPs. When VGG is deployed on mobile devices, it takes about 16 seconds to recognize a picture. Such a high inference latency is unacceptable in practical applications.

\begin{table*}[h]
\begin{center}
\caption{Popular DNN models}
\label{tab1}
\begin{tabular}{llrrr}
\toprule
\textbf{Model} & \textbf{Type} & \textbf{Parameters} & \textbf{Model Size} & \textbf{GFLOPs}\\
\midrule
LeNet & CNN & 431,080 & 1.64 MB & 0.005\\
AlexNet & CNN & 60,965,224 & 233 MB & 0.7\\
GoogleNet & CNN & 6,998,552 & 27 MB & 1.6\\
VGG-16 & CNN & 138,357,544 & 528 MB & 15.5\\
VGG-19 & CNN & 143,667,240 & 548 MB & 19.6\\
ResNet50 & CNN & 25,610,269 & 98 MB & 3.9\\
ResNet101 & CNN & 44,654,608 & 170 MB & 7.6\\
ResNet152 & CNN & 60,344,387 & 230 MB & 11.3\\
MobileNetV1 & CNN & 4,209,088 & 16 MB & 0.569\\
MobileNetV2 & CNN & 3,504,872 & 13 MB & 0.3\\
TextCNN & CNN & 151,690 & 0.6 MB & 0.009\\
YOLOV5s & CNN &7,266,973 &27.6 MB & 6.38 \\
Eng Acoustic Model & RNN & 34,678,784 & 132 MB & 0.035\\
RNNCell & RNN & 0.35 M & 1.33 MB & 0.01\\
BiRNN &RNN & 0.69 M &2.62 MB & 2.21\\
GRUCell & RNN & 1.04 M & 3.95 MB & 0.03\\
BiGRU &RNN & 2.07 M & 7.87 MB&6.65 \\
LSTMCell & RNN & 1.38 M & 5.24 MB & 0.04\\
BiLSTM &RNN &2.76 M & 10.49 MB& 8.86\\
\botrule
\end{tabular}
\end{center}
\end{table*}

\begin{table*}[h]
\centering
\caption{Popular DL hardware specifications}
\label{tab2}
\resizebox{\textwidth}{!}{
\begin{tabular}{lcrrr}
\toprule
\textbf{Equipment} & \textbf{Type} & \textbf{AI Performance} & \textbf{Memory} & \textbf{Bandwidth}\\
\midrule
V100 & GPU & 112 TFLOPS & 32 GB & 900 GB/s \\

A100  & GPU & 78 TFLOPS & 40 GB & 1555 GB/s \\

RTX 3090 & GPU & 35.58 TFLOPS & 24 GB & 936 GB/s \\

Titan RTX  & GPU & 32.62 TFLOPS & 24 GB & 672 GB/s \\

RTX 3080 & GPU & 29.77 TFLOPS & 10 GB & 760 GB/s \\

RTX 2080 Ti & GPU & 26.9 TFLOPS & 11 GB & 616 GB/s \\

Jetson AGX Xavier & Edge GPU & 32 TOPS & 32 GB & 136.5 GB/s \\

Jetson Xavier NX & Edge GPU & 21 TOPS & 8 GB & 51.2 GB/s \\

Jetson TX2 4GB & Edge GPU & 1.33 TFLOPS & 4 GB & 51.2 GB/s \\

Jetson TX2 &Edge GPU & 1.33 TFLOPS & 8 GB & 59.7 GB/s \\

Jetson TX2i & Edge GPU & 1.26 TFLOPS & 8 GB & 51.2 GB/s \\

Jetson TX1 & Edge GPU & 1 TFLOPS & 4 GB & 25.6 GB/s \\

Jetson Nano &Edge GPU & 0.47 TFLOPS  & 4 GB & 25.6 GB/s \\

Edge TPU & ASIC & 4 TFLOPS & - & - \\

Raspberry Pi-4B & ASIC & 13.5 GFLOPS & 4 GB & 8.5 GB/s \\

HONOR Magic3 & Mobile phone & 26 TOPS & 8 GB & 44 GB/s \\

iPhone 13 & Mobile phone & 15.8 TOPS & 4 GB & 34 GB/s \\

HUAWEI Mate40 & Mobile phone & - & 8 GB & 44 GB/s \\

Google Pixel6 & Mobile phone & - & 8 GB & 44 GB/s \\

Google Glass EE2 & Sensor & - & 3 GB &- \\
\botrule
\end{tabular}
\footnotetext{FLOPS: floating point operations per second, it can be understood as calculation speed. It is a measure of hardware performance.}
}
\end{table*}

As the core of AI, DNN relies on computing power and other resources of equipment. Table.~\ref{tab1} shows some specifications of popular hardware, including type, AI performance, memory and bandwidth, which determine the performance of the equipment in DNN inference. We can see that the single DNN inference mode of traditional AI has encountered a bottleneck. \cite{dinh2013survey}.

There are many problems in the machine learning system on a single terminal device \cite{gobieski2019intelligence}. Non-mobile terminals are usually free of computing resources, however, due to their non-mobile characteristics, they can not firstly obtain raw data, and the application scenario is greatly limited. Mobile devices generally have the limitations of battery capacity and computing power, it is impossible to deploy large-scale neural network models. However, just using a simple machine learning model will lead to the decline of system accuracy and affect the efficiency of task execution.

As for the traditional computing framework of IoT that completely relies on the cloud data for data processing, due to its strong computing power, the neural network model is usually deployed on the cloud server, and the training and inference tasks of the model are carried out on the cloud. In a cloud centric centralized IoT system, the IoT edge sensor is only responsible for collecting or generating data, sending the raw data to the cloud for processing, and sending the results back to the IoT edge device after inference. However, this cloud centric approach uploads a large amount of data to the remote cloud through long-time WAN data transmission, resulting in large end-to-end  transmission latency of mobile devices, high communication energy consumption, and high bandwidth occupancy. Furthermore, this kind of transmission may lead to data security and privacy problems \cite{ryan2011cloud}.

In order to solve the limitations of single terminal device method and the latency bottleneck of cloud centric method, we consider using the emerging distributed and cooperative computing method. Specifically, by assigning the inference and computing tasks from the core of network to the edge devices close to the terminal for collaborative execution, edge computing can realize DNN inference with low latency and energy saving. The hierarchical distributed collaborative computing structure composed of cloud, edge and terminal devices has inherent advantages \cite{skala2015scalable}, such as supporting coordinated central and local decision-making, and enhancing system scalability, which can be used for large-scale AI tasks of IoT devices based on geographic distribution.

\begin{figure}[h]
\centerline{\includegraphics[width=\textwidth]{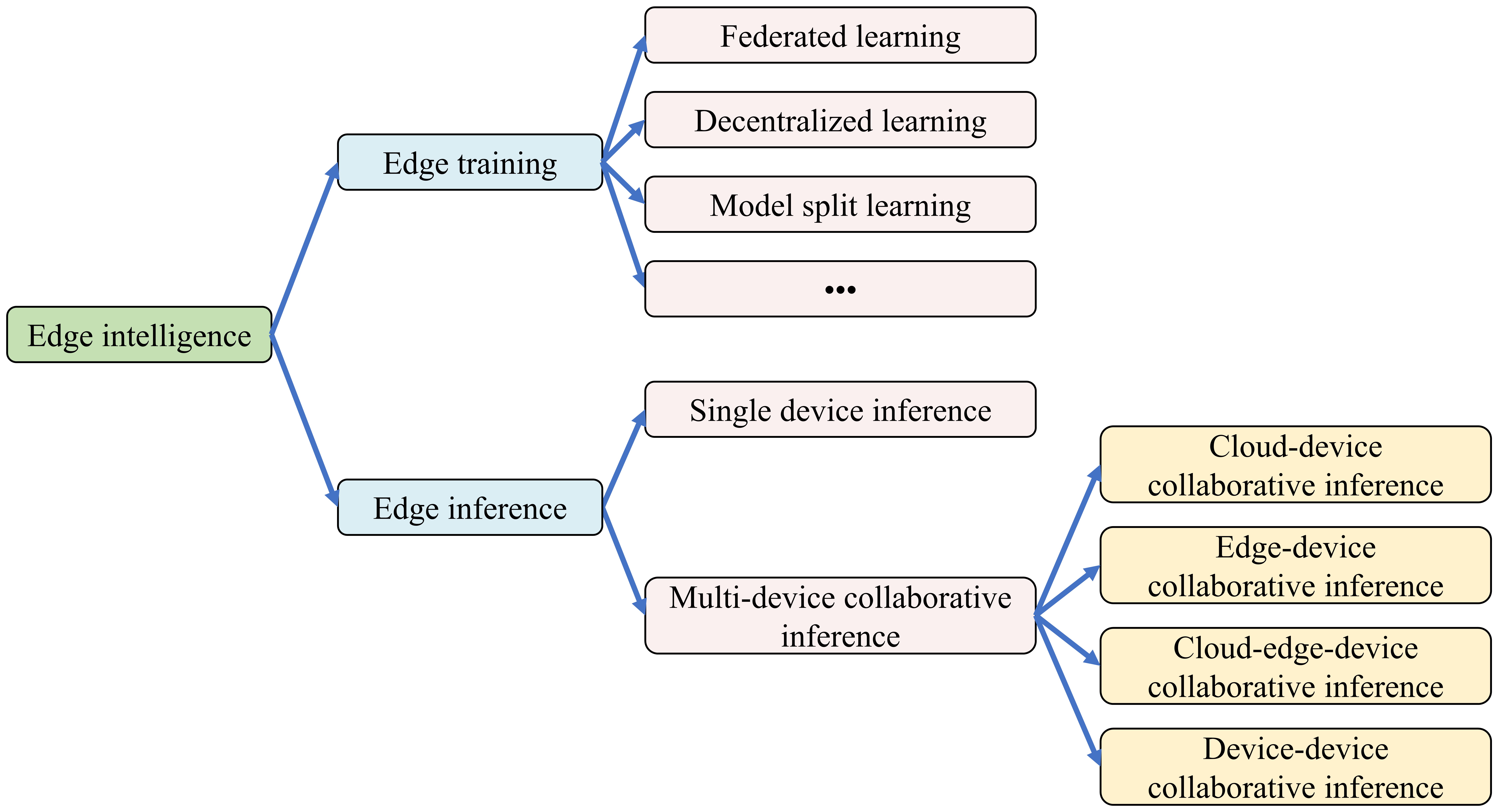}}
\caption{Classification of EI.\label{fig2}}
\end{figure}

\subsection{Various collaborative DNN inference paradigms}
Deep learning represents the most advanced AI technology, and it is naturally suitable for edge computing. At the same time, the collaborative DNN inference paradigm can make full use of the available data and resources in the hierarchical structure of terminal equipment, edge nodes and cloud data center to optimize the overall performance of training and inference of DNN model. This shows that EI oriented collaborative DNN inference does not necessarily mean that DNN model infers on a single device or cloud, but it can work in the way of cloud edge device coordination through data offloading. Specifically, according to the way of task offloading and path length, we divide the collaborative DNN inference into four collaborative DNN inference paradigms. As shown in Fig.~\ref{fig2}, the definitions of various collaborative DNN inference paradigms are as follows:
\begin{enumerate}[1)]
\item \textbf{Cloud-device collaborative DNN inference}. The DNN model is deployed on cloud server and terminal devices, and DNN inference is carried out through the cooperation between cloud and terminal devices. The terminal devices will partially process the raw data and transfer the extracted feature information data to the cloud. The cloud will receive the remaining data for inference and send the final decision result back to the terminal device. \emph{The cloud-device collaborative DNN inference paradigm will pay more attention to latency and it is usually used in scenarios with weak mobility.}
\item \textbf{Edge-device collaborative DNN inference}. The DNN model is deployed on the edge server and terminal devices, and the DNN model is inferred through the cooperation between the edge and terminal devices. Model inference is performed within the network edge, which can be achieved by offloading all or part of the data to the edge nodes or nearby devices. \emph{This paradigm focuses on inference accuracy and it can be employed in highly interactive application scenarios.}
\item \textbf{Cloud-edge-device collaborative DNN inference}. The DNN model is deployed on cloud servers, edge servers and terminal devices at the same time, and the DNN model is deduced through the cooperation of the three. \emph{This paradigm focuses on total cost and stability. It can be used in scenarios with large amount of calculation and data.}
\item \textbf{Device-device collaborative DNN inference}. Deploy the DNN model on the local terminal devices, and conduct DNN inference completely in the way of local cooperation. It means that the data will be processed near the source to obtain the decision results without the participation of edge servers and cloud servers. \emph{This paradigm focuses on inference latency and energy consumption, but it can be applied in high mobility scenarios or some remote and harsh environments.}
\end{enumerate}

As the center of collaborative DNN inference shifts to the edge, the amount of offloaded data and path length decrease gradually. Therefore, it gets lower transmission latency of data offloading, lower cost of communication bandwidth, and higher data security. However, this is at the cost of increasing computing latency and energy consumption. This conflict shows that the optimal collaborative DNN inference paradigm depends on the application programs and application scenarios, and should be determined by jointly considering multiple standards, such as latency, energy efficiency, privacy, and bandwidth cost. In later sections, we will review different collaborative DNN inference paradigms and existing solutions.

\section{Cloud-device collaborative DNN inference}\label{sec3}
In this section, we firstly introduce the motivation and architecture of cloud-device collaborative DNN inference paradigm. Then We summarize the performance optimization of cloud device collaborative DNN inference in recent years. Finally, we compare the differences between cloud-device collaborative DNN inference and traditional centralized cloud computing, and analyze the shortcomings that the proposed cloud-device collaborative DNN inference still exist.
\subsection{Motivation and architecture}
\begin{figure}[h]
\centerline{\includegraphics[width=\textwidth]{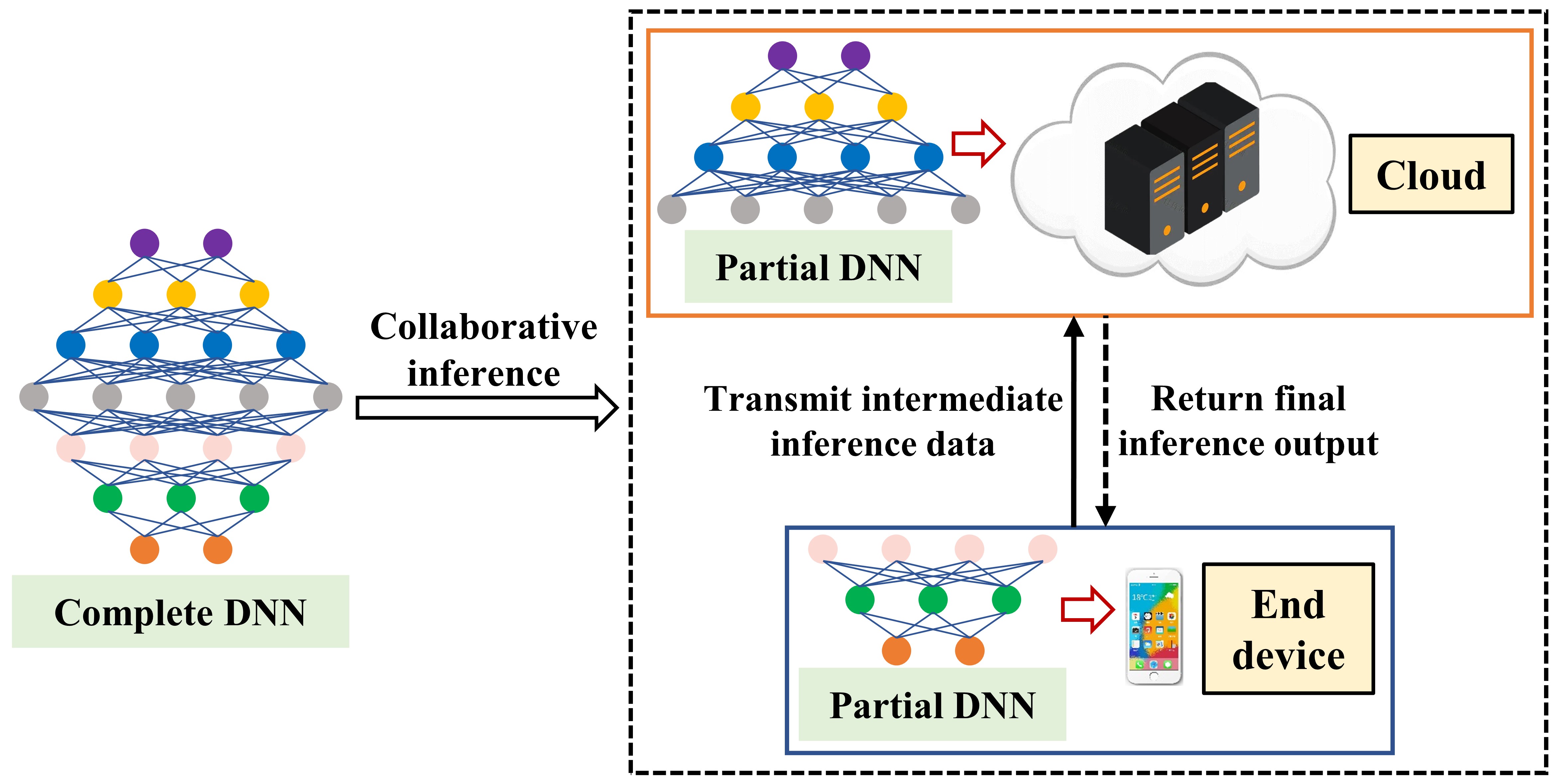}}
\caption{Illustration of cloud-device collaborative DNN inference.\label{fig3}}
\end{figure}

The traditional cloud-only DNN inference computing method needs to upload a large amount of data, such as image, video and audio, to the server through wireless network, resulting in high latency and energy cost. Therefore, data transmission becomes a bottleneck in traditional cloud-only DNN inference. As shown in Fig.~\ref{fig3}, different from the traditional centralized cloud computing framework, cloud-device collaborative DNN inference makes more precise task scheduling and allocation of terminal devices and cloud servers, and strengthens the collaborative operation mode. More specifically, we can calculate a part of DNN on the edge side, transfer a small amount of intermediate results to the cloud, and then calculate the rest in the cloud. The division of DNN constitutes a trade-off between calculation and transmission. \emph{However, due to the limitations of cloud including the equipment that cannot be moved, cloud-device collaborative inference only can be used in scenarios with weak mobility, such as mall monitoring or provide web service to users. Considering the application scenarios of cloud-device collaborative inference, the performance optimization will pay more attention to the latency, because the transmission time is the main problem in weak mobility scenarios.}

\subsection{Performance optimization}
According to different objectives of performance optimization, cloud-device collaborative DNN inference can be divided into two categories. And as shown in Table.~\ref{tab3}, we also collect application, main technology and effectiveness of each cloud-device collaborative inference framework.

\begin{table*}[h]
\centering
\caption{Classification of cloud-device collaborative DNN inference}
\label{tab3}
\resizebox{\textwidth}{!}{
\begin{tabular}{ccccc}
\toprule
\textbf{Optimization Objective} &\textbf{Framework} & \textbf{Application} & \textbf{Main Technology} & \textbf{Effectiveness} \\
\midrule
\multirow{8}{1.5in}{Total inference latency minimization}
&Split-brain \cite{emmons2019cracking} & Video analysis & Model partition & \makecell[l]{Communication compression: 2-3$\times$}\\
\cline{2-5}
&In-situ AI \cite{song2018situ} & IoT & Model partition & \makecell[l]{Data movement reduction: 28-71$\%$ \\ Model update acceleration: 1.4-3.2$\times$ \\ Energy consumption reduction: 30-70$\%$} \\
\cline{2-5}
&DADS \cite{hu2019dynamic} & Video analysis & Model partition & \makecell[l]{Latency reduction: 6.45-8.08$\times$ \\ Throughput improvement: 8.31-14.01$\times$} \\
\cline{2-5}
&DeepInference-L \cite{wang2021accelerate} & IoT & Model partition & \makecell[l]{Inference speedup: 8$\times$} \\
\cline{2-5}
&IONN \cite{jeong2018ionn} & IoT &Model partition &\makecell[l]{Lower latency} \\
\midrule
\multirow{12}{1.5in}{Energy consumption minimization}
&Neurosurgeon \cite{kang2017neurosurgeon} & Computer vision & Model partition & \makecell[l]{Latency reduction: 3.1$\times$ \\ Energy consumption reduction: 59.5$\%$ \\ Throughput improvement: 1.5$\times$}\\
\cline{2-5}
&Vision Pipeline \cite{hauswald2014hybrid} & Computer vision & \makecell[c]{Calculation offloading\\ Image compression} & \makecell[l]{Data transmission reduction: 2.5$\times$\\ Energy consumption reduction: 3.7$\times$}\\
\cline{2-5}
&\cite{laskaridis2020spinn} &Computer vision &\makecell[c]{Model partition\\ Early-exit} &\makecell[l]{Throughput improvement: 2$\times$\\Energy consumption reduction: 6.8$\times$\\Accuracy improvement: 20.7$\%$} \\
\cline{2-5}
&JointDNN \cite{eshratifar2019jointdnn} & Mobile intelligence & Model partition & \makecell[l]{Latency reduction: 18$\times$\\ Energy consumption reduction: 32$\times$} \\
\cline{2-5}
&\cite{deng2016fine}&IoT &\makecell[c]{Model partition\\Calculation offloading} &\makecell[l]{Energy consumption reduction: 25$\%$} \\
\botrule
\end{tabular}
}
\end{table*}

\subsubsection{Total inference latency minimization}
In the applications of computer vision and video analysis, real-time is generally the focus of attention. For example, in the automatic driving vehicle application \cite{gerla2014internet}, the camera continuously monitors the surrounding scene and transfers it to the server, then the server performs video analysis, and sends the control signal to the pedals and steering wheels in real-time to cope with the environmental changes. In the augmented reality applications, the intelligent device will continuously record the current view and transmit the information to the cloud server, which will perform object recognition and send back the enhancement label for real-time rendering on the actual scene. A major obstacle to intelligent applications is the large amount of data in video streams. Therefore, how to reduce the total inference latency is an important research point of cloud-device collaborative DNN inference.

At present, DNN has made rapid progress in system design, but the existing systems still regard DNN as a ``black box". Whether the model is fully deployed on the terminal device or the video is compressed into the cloud for analysis, these two methods will affect the accuracy and total cost of inference. John Emmons \emph{et al.} \cite{emmons2019cracking} propose to open the black box of neural network and describe new application scenarios to enable collaborative inference between terminal IoT devices and cloud.

The best partition point of a DNN architecture depends on the topology of DNN, which is reflected in the change of calculation and data size of each layer. The division of different layers will lead to different computing latency and transmission latency. Therefore, an optimal partition is needed. In addition, even under the same DNN architecture, dynamic factors such as wireless network status and data center load will affect the optimal partition point.

Considering that the terminal devices in the traditional centralized cloud computing framework will collect or generate a large amount of raw data for transmission, song \emph{et al.} \cite{song2018situ} propose to improve the accuracy of the IoT system with minimal data movement, and designed an autonomous incremental computing framework and architecture named in-situ AI for IoT applications based on deep learning. It is used to solve the problem of labeling IoT data and release the raw data potential of the IoT system to reduce the latency of training and inference.

The latest progress of DNN shows that DNN is no longer limited to chain topology, and directed acyclic graph (DAG) topology is becoming more and more popular. For example, GoogleNet \cite{szegedy2015going} was the champion of the ImageNet in 2014 and ResNet \cite{he2016deep} was the champions in 2015, both of which are DAG topology. However, compared with the chain structure DNN, the dividing of DAG topology needs more complex analysis of the model, which may lead to Non-deterministic Polynomial (NP) hard problems in performance optimization.

Based on the fact that the data size of some intermediate DNN layers is smaller than the raw input data, the optimal DNN division can be found in the integrated cloud computing environment with dynamic network conditions. Hu \emph{et al.} \cite{hu2019dynamic} design a dynamic adaptive DNN division scheme, which allows partition DNN to be processed on the terminal device and cloud, while limiting data transmission. The scheme optimizes the division of DNN network by continuously monitoring the network status. During continuous network monitoring, the scheme determines whether the system under light load or heavy load. If under light load conditions, the system will minimize the total latency of processing a frame. While if under heavy load, it will maximize the throughput which means the number of frames that can be processed per unit time.

In order to reduce the communication latency introduced in computing offloading, Wang \emph{et al.} \cite{wang2021accelerate} propose collaborative deep inference, which divided the DNN model into two parts. In this method, DNN calculation and communication are carried out at the same time, and it is proposed that the optimal partition can be found in the DNN of DAG computing architecture. Jeong \emph{et al.} \cite{jeong2018ionn} propose \emph{IONN}, a DNN offloading technology based model partition to reduce inference latency and improve query performance.

\subsubsection{Energy consumption minimization}
In cloud-device collaborative DNN inference, since the battery capacity of mobile devices in some application scenarios will also be limited, minimizing energy consumption is also an aspect we need to consider. Of course, cloud servers account for a large part of inference computing in cloud-device collaborative DNN inference paradigm. Compared with single terminal device, the energy consumption limit in inference scenarios is not very obvious, so such inference paradigm has relatively little research on energy consumption.

Kang \emph{et al.} \cite{kang2017neurosurgeon} research the computing partition strategies based on the traditional cloud-only processing methods. These strategies can effectively use the cloud and mobile devices to realize intelligent applications with low latency, low energy consumption and high data center throughput. The authors also design a lightweight dynamic scheduler \emph{neurosurgeon}, which can automatically divide DNN computing between mobile devices and cloud servers. \emph{Neurosurgeon} is a runtime system that spans the cloud and mobile platforms. It can automatically identify the best partition points in DNN and coordinate the allocation of computing tasks by using the processing capacity of mobile devices and cloud servers, rather than completely limiting the inference computing to the cloud or terminal, which reduces the communication energy consumption of data transmission and realizes far more latency performance and energy efficiency.

Hauswald \emph{et al.} \cite{hauswald2014hybrid} study the trade-offs between mobile devices and cloud server when performing part of the workload. They analyze the ability of mobile devices to perform feature extraction and prediction under optimal configuration. Through the preliminary processing of data by mobile devices, lower data transmission and energy consumption can be realized. Laskaridis \emph{et al.} \cite{laskaridis2020spinn} propose a distributed inference system called \emph{SPINN}, which employs collaborative cloud-device computing and progressive inference methods to provide fast and stable CNN inference in different environments. Compared with cloud-only mode, it significantly improves throughput and reduces energy consumption.

In paper \cite{eshratifar2019jointdnn}, Eshratifar \emph{et al.} propose an efficient, adaptive and practical engine, \emph{JointDNN}, for collaborative inference between mobile devices and cloud. \emph{JointDnn} not only provides a performance efficient method for collaborative DNN, but also reduces the workload and communications compared with cloud-only method. Deng \emph{et al.} \cite{deng2016fine} propose a fine-grained offloading strategy to minimize energy consumption while meeting strict latency constraints.

\subsection{Summary and analysis}
Cloud-device collaborative DNN inference alleviates some disadvantages of the traditional cloud-only inference computing method to a certain extent:
\begin{enumerate}[1)]
\item \textbf{Inference latency reduction}. Cloud-device collaborative DNN inference divides the DNN model and assigns the inference computing task to terminal devices and cloud server for collaborative execution which reduces the inference latency.
\item \textbf{Energy consumption reduction}. Since cloud-device collaborative DNN inference reduces the communication cost by model partition and task allocation, it reduces the energy consumption of computing and communication.
\end{enumerate}

However, the cloud-device collaborative DNN inference paradigm still has some shortcomings:
\begin{enumerate}[1)]
\item \textbf{Insufficient performance under mobile and harsh scenarios}. Cloud-device collaborative DNN inference cannot meet the real-time and high-precision requirements of image processing in high-speed mobile scenes and harsh communication environments.
\item \textbf{Limited application scenarios}. Since the cloud server is far away from the terminal device, and it can not remove, the application scenarios of cloud-device collaborative DNN inference are limited.
\end{enumerate}

\section{Edge-device collaborative DNN inference}\label{sec4}
In this section, we firstly introduce the motivation and architecture of edge-device collaborative DNN inference paradigm. Then we discuss the performance optimization of the researches in recent years. Finally, we summarize the advantages of edge-device collaborative DNN inference, and analyze the shortcomings that still exist.

\subsection{Motivation and architecture}
\begin{figure}
\centerline{\includegraphics[width=\textwidth]{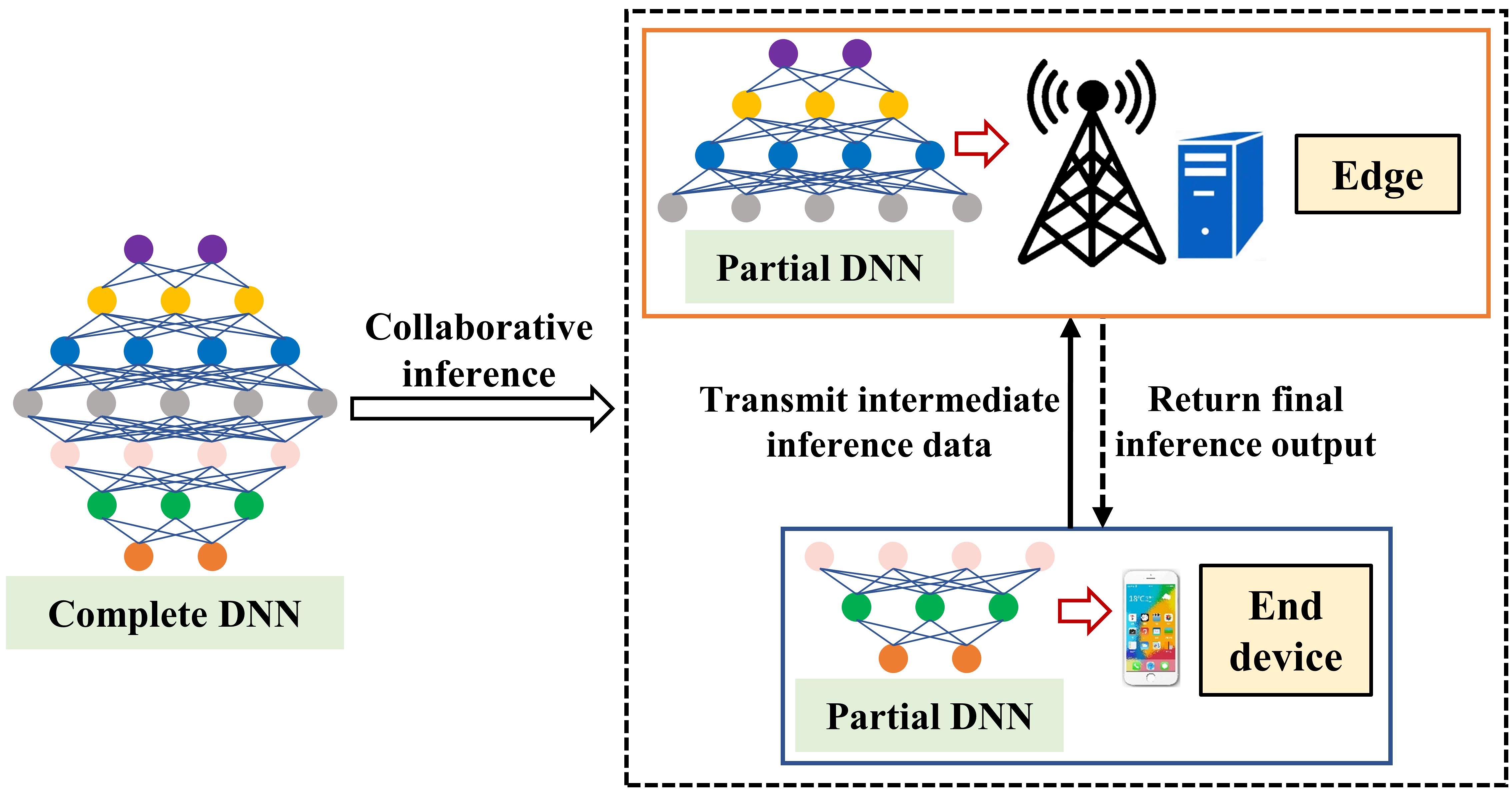}}
\caption{Illustration of edge-device collaborative DNN inference.\label{fig4}}
\end{figure}

In order to alleviate the latency and energy bottleneck, the emerging edge computing paradigm has been introduced into the intelligent industry. As the computing center continues to shift to the edge of the network, how to achieve low latency and energy saving inference task processing is what we need to consider. Although cloud-device collaborative DNN inference solves the problems of high latency and high bandwidth consumption caused by mass of data transfer of the traditional centralized cloud processing to a certain extent, it also has some limitations like the centralized cloud processing framework. Considering that the network edge is becoming another potential collaborator, it cannot only reduce the computing load of the central site in a distributed way, but also improve the service performance due to the close distance to the users. In order to further reduce the communication consumption and inference latency, many scholars have proposed edge-device collaborative DNN inference paradigm. In Fig.~\ref{fig4}, it exhibits the illustration of edge-device collaborative inference. The edge server instead of the cloud server reduces the distance between terminal devices and severs. \emph{The collaborative paradigm can deal with real-time tasks in highly interactive scenarios because the transmission distance is shorter. In addition, it gets a technology called early-exit to improve the performance. However, early-exit technology will reduce the inference accuracy to a certain extent, thus the performance optimization of edge-device collaborative inference should pay more attention to the inference accuracy.}

\subsection{Performance optimization}
According to different performance optimization objectives, we can divide edge-device collaborative DNN inference into three categories, as shown in Table.~\ref{tab4}.

\begin{table*}[h]
\caption{Classification of edge-device collaborative DNN inference}
\label{tab4}
\centering
\resizebox{\textwidth}{!}{
\begin{tabular}{ccccc}
\toprule
\textbf{Optimization Objective} & \textbf{Framework} & \textbf{Application} & \textbf{Main Technology} & \textbf{Effectiveness} \\
\midrule
\multirow{4}{1.5in}{Total inference latency minimization}
&DINA \cite{mohammed2020distributed} & IoT & \makecell[c]{Model partition\\Data partition} & \makecell[l]{Latency reduction: 2.6-4.2$\times$}\\
\cline{2-5}
&Cogent \cite{shan2020collaborative} & Intelligence service & \makecell[c]{Model compression\\Model partition} & \makecell[l]{Inference acceleration: 8.89$\times$}\\
\cline{2-5}
&CoopAI \cite{yang2021cooperative} & IoT &\makecell[c]{Model partition} &\makecell[l]{Inference acceleration: 20-30$\%$} \\
\cline{2-5}
&SCADS \cite{liu2018scads} & Mobile intelligence & \makecell[c]{Partial offloading}& \makecell[l]{Latency reduction: 58$\%$} \\
\midrule
\multirow{15}{1.5in}{Inference accuracy maximization}
&$O^3$ \cite{hanyao2021edge} & Video analysis & \makecell[c]{Edge assistance} & \makecell[l]{Accuracy improvement: 36$\%$ }\\
\cline{2-5}
&\cite{yun2021cooperative} & Image processing & \makecell[c]{Model partition} & \makecell[l]{Maximizing the accuracy\\ while satisfying the\\ latency constraint}\\
\cline{2-5}
&Edgent \cite{li2018edge,li2019edge} & Mobile intelligence &\makecell[c]{Model partition\\ Model right-sizing\\ Early-exit} & \makecell[l]{Maximizing the accuracy\\ while satisfying the\\ latency constraint}\\
\cline{2-5}
&\cite{song2021adaptive} & IoT & \makecell[c]{Model partition\\ Early-exit} & \makecell[l]{Maximizing the accuracy\\ while satisfying the\\ latency constraint}\\
\cline{2-5}
&Boomerang \cite{zeng2019boomerang} & IIoT & \makecell[c]{Model partition\\ Model right-sizing \\ Early-exit} & \makecell[l]{Lower latency\\ Higher accuracy}\\
\cline{2-5}
&PADCS \cite{hu2021enable} & Video analysis &\makecell[c]{Model partition\\ Early-exit\\ Data compression} & \makecell[l]{Performance improvement: 1.5-2.8$\times$\\Higher stability} \\
\midrule
\multirow{9}{1.5in}{Total cost minimization}
&NestDNN \cite{fang2018nestdnn} & Mobile vision & \makecell[c]{Dynamic resource\\Model compression\\Model recovery} & \makecell[l]{Accuracy improvement: 4.2$\%$\\Efficiency improvement: 2$\times$\\Eenergy consumption reduction: 1.7$\times$}\\
\cline{2-5}
&CORA \cite{du2017computation} & Mobile computing &\makecell[c]{Resource allocation\\ Computation offloading} &\makecell[l]{Lower cost}\\
\cline{2-5}
&\cite{tang2020joint} & IoT &\makecell[c]{Model partition\\Resource allocation} &\makecell[l]{Resource reduction: 63$\%$\\Efficiency improvement: 41-67.6$\%$}\\
\cline{2-5}
&HMTD \cite{yang2020offloading} & Mobile vision &\makecell[c]{Model partition} &\makecell[l]{Latency minimization\\ Energy consumption minimization}\\
\cline{2-5}
&\cite{dong2021joint} &IoT &\makecell[c]{Model partition\\ Resource allocation} &\makecell[l]{Rental cost reduction: 30$\%$}\\
\botrule
\end{tabular}
}
\end{table*}

\subsubsection{Total inference latency minimization}
In order to overcome the related resource constraints, inference tasks are offloaded to the edge or cloud through DNN partition. However, most existing solutions only divide DNN into two parts, one running locally and the other running in the cloud. In contrast, Mohammed \emph{et al.} \cite{mohammed2020distributed} propose a technology called \emph{DINA} to divide DNN into multiple partitions, which can be processed locally by terminal device or offloaded to one or more powerful nodes. The authors creatively combine the matching theory \cite{roth1992two} with DNN inference task offloading, and reduce the amount of computation through adaptive DNN partition and distributed algorithm based on matching theory, so as to significantly reduce the total latency of DNN inference. Yang and Kuo \emph{et al.} \cite{yang2021cooperative} design a collaborative edge computing system called \emph{CoopAI} to minimize the inference latency. It distributes DNN inference on multiple edge devices through a new model partition technology, allowing edge devices to preload the required data in advance, so as to calculate inference in parallel without exchanging data.

Shan \emph{et al.} \cite{shan2020collaborative} propose \emph{cogent}, which accelerates deep neural network inference through edge-device cooperation. \emph{Cogent} includes automatic pruning phase and container deployment phase. The pruning and partitioning model can better adapt to the system environment and hardware configuration. The flexibility and reliability of the system can be improved by deploying dynamic packaging modules in containers and assigning tasks to terminal devices and edge servers. \emph{Cogent} makes full use of edge devices for collaborative inference, which significantly reduces service latency while maintaining accuracy. Liu and Haoran \emph{et al.} \cite{liu2018scads} study the latency optimization problem of mobile edge-device collaborative inference, and propose a simultaneous calculation and allocation strategy for task offloading.

\subsubsection{Inference accuracy maximization}
Since the computing power of edge server is inferior to remote cloud server, compared with cloud-device collaborative DNN inference, edge-device collaborative DNN inference not only needs to pay attention to the efficiency of inference, but also needs to improve the accuracy of inference.

Considering that it is inefficient to transfer the target detection task to the edge, Hanyao \emph{et al.} \cite{hanyao2021edge} propose a system with the functions of target tracking on the terminal device and auxiliary analysis on the edge at the same time. Through the cooperative inference of edge end devices, the overall accuracy of target detection can be improved to the greatest extent under the conditions of dynamic edge network and limited detection latency.

Yun \emph{et al.} \cite{yun2021cooperative} consider the actual noisy wireless channel between the device and edge server. Therefore, an automatic repeat request method and a practical error correction code are adopted in the cooperative DNN inference to ensure the latency and improve the accuracy at the same time.

\begin{figure}
\centerline{\includegraphics[width=\textwidth]{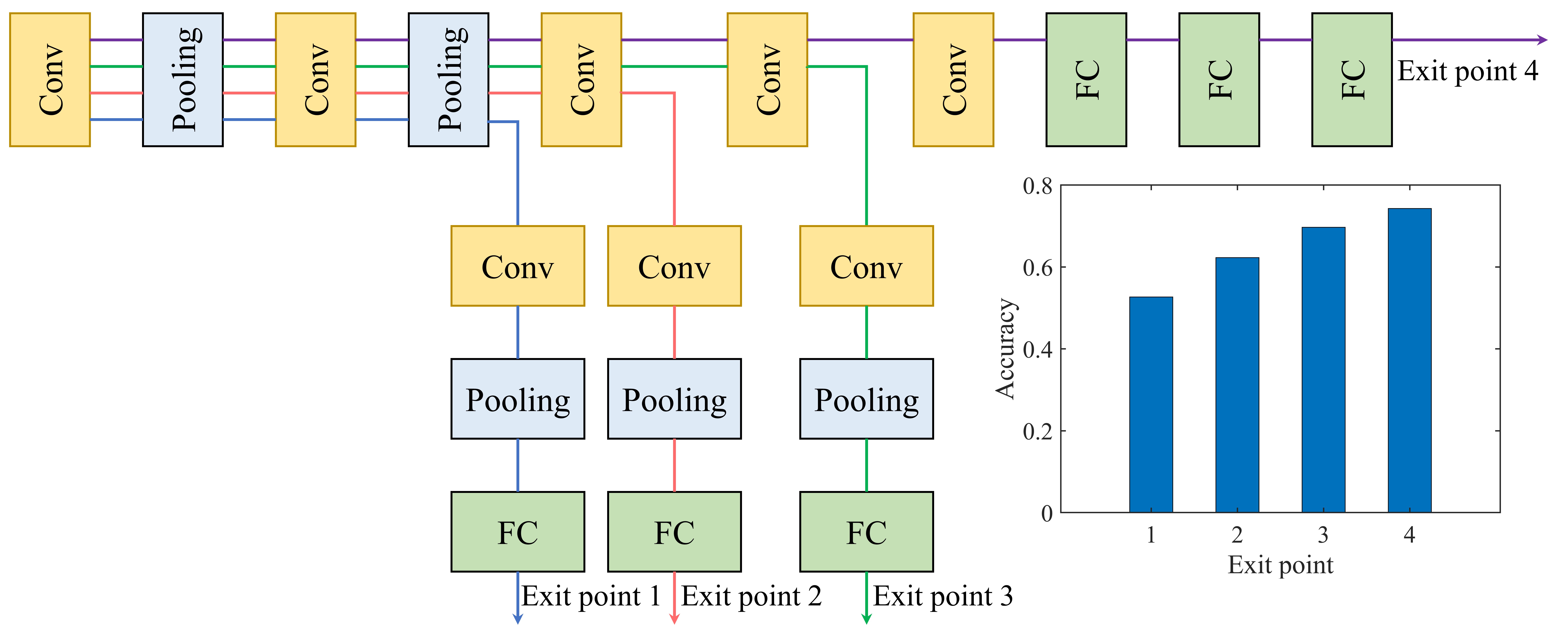}}
\caption{Illustration of early-exit mechanism in branchy AlexNet model.~\cite{teerapittayanon2016branchynet}\label{fig5}}
\end{figure}

Teerapittayanon \emph{et al.} \cite{teerapittayanon2016branchynet} propose \emph{BranchyNet}, which first employs early-exit mechanism. As shown in Fig.~\ref{fig5}, the network exists several exit points and it can select exit point according to performance requirements. From the perspective of framework design, many researches of edge-device collaborative DNN inference add early-exit mechanism on the basis of DNN division. EI realized by early-exit mechanism effectively reduces the waste of storage resources by adjusting the size of the model, and is more suitable for deployment on resource-limited devices. However, because the method of adjusting the size of DNN model will reduce the accuracy of inference to a certain extent, many studies using early-exit mechanism focus on improving the accuracy of inference within the allowable range of latency.

Li \emph{et al.} \cite{li2018edge,li2019edge} propose a deep learning model collaborative inference framework based on edge-device collaboration, which named \emph{edgent}. For EI with low latency, edge combines DNN partitioning and early-exit mechanism to reduce the latency of inference tasks. To improve the performance, \emph{edgent} jointly optimizes DNN partitions and resizes on demand. For tasks with deadlines, \emph{edgent} can maximize accuracy without exceeding the deadline. Song \emph{et al.} \cite{song2021adaptive} also propose a collaborative inference system based on \emph{edgent}, which applies early-exit mechanism and model division technology to solve the problem of EI in task flow scenarios. At the same time, the authors design an offline dynamic programming algorithm and an online deep reinforcement learning algorithm to dynamically select the exit point and partition point of the model in the task flow, so as to balance the efficiency and accuracy of inference tasks.

In order to realize the real-time industrial application based on DNN in the edge computing paradigm, Zeng \emph{et al.} \cite{zeng2019boomerang} propose \emph{boomerang}, which is an on-demand collaborative DNN inference framework for EI in IoT environment. \emph{Boomerang} uses DNN partition and early-exit mechanism to perform DNN inference tasks with low latency and high precision. Hu \emph{et al.} \cite{hu2021enable} describe the acceleration problem of multiple collaborative inference tasks as a pipeline execution model, and design a fine-grained optimizer, which integrates model partition, early model exit and intermediate data compression to achieve a trade-off between accuracy and latency.

\subsubsection{Total cost minimization}
Due to the limited resources of edge devices, how to minimize the computing overhead and equipment rental cost must be considered in the edge-device collaborative inference. Fang \emph{et al.} \cite{fang2018nestdnn} adopt the idea of DNN filter pruning to adjust the size of the dynamic model to reduce the total computational offloading cost. However, filter pruning will reduce the inference accuracy, it is difficult to achieve a good balance between cost reduction and accuracy loss. Du \emph{et al.} \cite{du2017computation} discuss the joint resource management of multiple equipments, but only support a fixed number of offload decisions. Tang \emph{et al.} \cite{tang2020joint} study the optimization problem of DNN partition under realistic multi-user resource constraints, revealed some properties of the optimization problem of multi-user DNN joint partition and computing resource allocation, and balanced the task inference accuracy and computing overhead.

Unmanned aerial vehicle (UAV) is widely used in target tracking and other applications, but it is difficult to complete the tasks requiring intensive computing independently due to the serious limitation of UAV power supply and low computing power. Therefore, based on the limited computing resources and the energy budget of UAV, Yang \emph{et al.} \cite{yang2020offloading} propose a new layered task allocation framework, in which UAV is embedded in the lower layer of pre training CNN model, and the mobile edge computing server with rich computing resources will process the higher layer of CNN model. The authors also propose an optimization problem to minimize the weighted sum cost, including the tracking latency and energy consumption introduced by UAV communication and calculation, while considering the data quality and inference error.

Due to the shortage of edge computing resources and high rental cost, it is more difficult to optimize the task execution based on DNN. Dong \emph{et al.} \cite{dong2021joint} propose a joint method of adaptive DNN partition and cost-effective resource allocation, which balances the inference latency and overall rental cost of DNN tasks, so as to promote the collaborative computing between IoT devices and edge servers.

\subsection{Summary and analysis}
Compared with cloud-only mode and cloud-device mode, edge-device collaborative DNN inference gets several advantages as follows:
\begin{enumerate}[1)]
\item \textbf{Lower latency}. Edge-device collaborative inference gets lower latency because it reduces the communication distance to achieve lower communication latency, and employs early-exit mechanism to resize inference model which can reduce the computing latency.
\item \textbf{Lower cost}. Since the communication distance becomes shorter, the communication cost is lower. The usage of early-exit mechanism reduces the waste of storage resources and computing cost. In addition, edge-device collaborative inference employs edge equipment as the server, which saves the rental cost.
\end{enumerate}

Although edge-device collaborative DNN inference achieves many advantages, it still faces some challenges, such as the limitation of real-time performance, because the efficiency of edge based DNN inference is highly dependent on the available bandwidth between edge servers and IoT devices. With the sinking of the computing center, edge-device collaborative inference paradigm may be considered as an extension of cloud-device collaborative inference paradigm.

\section{Cloud-edge-device collaborative DNN inference}\label{sec5}
In this section, we firstly introduce the motivation and architecture of cloud-edge-device collaborative DNN inference paradigm. Then we discuss the performance optimization of the researches in recent years. Finally, we compare the differences between cloud-edge-device collaborative DNN inference paradigm and the previous two collaborative DNN inference paradigms, which include advantages and shortcomings.

\subsection{Motivation and architecture}
The offloading of DNN model can reduce the pressure of mobile devices by transferring intensive computing from resource constrained devices to cloud or edge servers, so as to accelerate DNN inference. Cloud-device collaborative DNN inference usually offloads some DNN models to cloud servers with high computing power, so as to reduce the inference latency of tasks. However, since the cloud center is too far away from the client, the performance of inference will be affected by factors such as network bandwidth, computing power, the amount of data transmitted and the number of computing tasks. Edge servers are widely distributed among mobile devices and cloud computing centers, and integrate the core functions of network, computing, storage and applications. Therefore, edge-device collaborative DNN inference combined with edge computing can effectively reduce the burden of network bandwidth and achieve lower transmission latency. However, the computing power of edge server is usually limited, so it is difficult to support the inference of large-scale DNN model.

\begin{figure}
\centerline{\includegraphics[width=\textwidth]{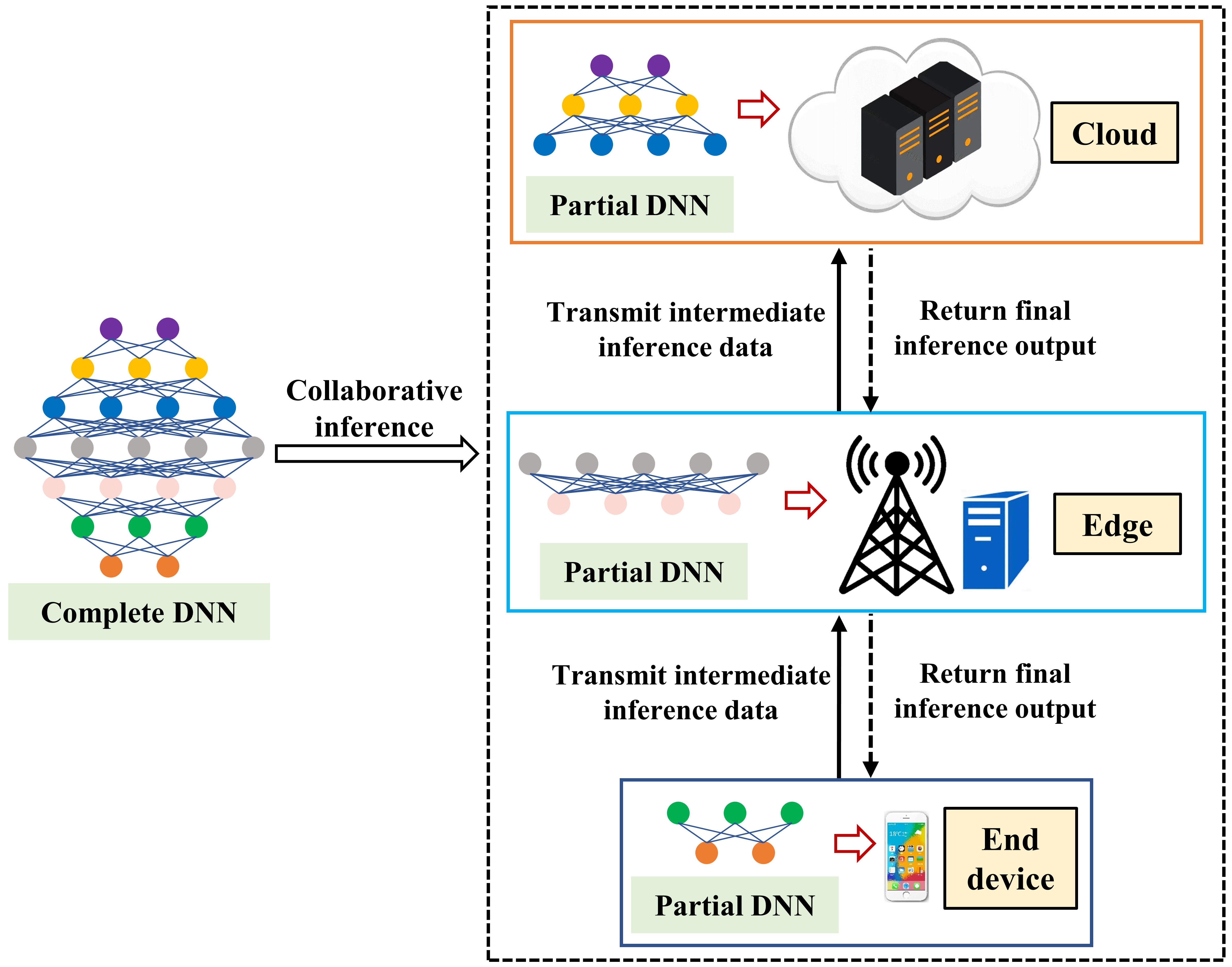}}
\caption{Illustration of cloud-edge-device collaborative DNN inference.\label{fig6}}
\end{figure}

Due to different scenario applications and different physical distances of collaborators, cloud-device collaborative DNN inference and edge-device collaborative DNN inference do not make full use of heterogeneous devices at all levels. As shown in Fig.~\ref{fig6}, cloud-edge-device collaborative DNN inference combines the characteristics of cloud computing with high computing power and edge computing with low transmission latency, enhances the interaction between devices and improves the flexibility and scalability of the system. At the same time, paper \cite{morshed2017deep} analyzes the challenges involved in developing cloud-edge-device distributed collaborative deep learning algorithms. These algorithms have resource and data awareness, and can consider the underlying heterogeneous data model, resource model and data availability when performing tasks. \emph{However, since the increase of participating equipment, performance optimization objectives of cloud-edge-device collaborative inference will consider total cost and stability of the system. This paradigm can be used in scenarios with large amount of calculation and data, meanwhile, a good communication environment is necessary.}

\subsection{Performance optimization}
According to different performance optimization objectives, we divide cloud-edge-device collaborative DNN inference into three categories, as shown in Table.~\ref{tab5}.

\begin{table*}[h]
\caption{Classification of cloud-edge-device collaborative DNN inference}
\label{tab5}
\centering
\resizebox{\textwidth}{!}{
\begin{tabular}{ccccc}
\toprule
\textbf{Optimization Objective} & \textbf{Framework} & \textbf{Application} & \textbf{Main Technology} & \textbf{Effectiveness} \\
\midrule
\multirow{6}{1.5in}{Total inference latency minimization}
&\cite{ren2021fine} & Mobile intelligence & \makecell[c]{Model partition} & \makecell[l]{Latency reduction: 47.02-72.17$\%$\\Throughput improvement: 0.875$\times$\\Energy consumption reduction: 66.91$\%$} \\
\cline{2-5}
&EosDNN \cite{xue2021eosdnn} & Computer vision & \makecell[c]{Model migration \\Model partition} & \makecell[l]{Lower latency} \\
\cline{2-5}
&STEED \cite{lin2019distributed} &- &\makecell[c]{Model partition} &\makecell[l]{Performance improvement: 2$\times$} \\
\cline{2-5}
&\cite{dey2019offloaded} &- & \makecell[c]{Model partition\\Intelligence offloading} &\makecell[l]{Latency reduction: 2$\times$} \\
\midrule
\multirow{8}{1.5in}{Total cost minimization}
&\cite{lin2019cost} &- & \makecell[c]{Model partition} &\makecell[l]{Lower system cost} \\
\cline{2-5}
&\cite{teerapittayanon2017distributed} &- & \makecell[c]{Model partition\\Early-exit} &\makecell[l]{Communication cost reduction: 20$\times$}\\
\cline{2-5}
&EdgeEye \cite{liu2018edgeeye} &Computer vision & \makecell[c]{Model partition} &\makecell[l]{Lower latency\\Lower cost\\Higher accuracy}\\
\cline{2-5}
&eSGD \cite{tao2018esgd} &IoT & \makecell[c]{Model partition} &\makecell[l]{Lower communication cost\\Higher accuracy} \\
\midrule
\multirow{3}{1.5in}{Failure-resilient distributed DNN model}
&deepFogGuard \cite{yousefpour2019guardians} &IoT & \makecell[c]{Model partition\\ Skip hyperconnections} &\makecell[l]{Fault recovery ability}\\
\cline{2-5}
&ResiliNet \cite{yousefpour2020resilinet} &IoT &\makecell[c]{Model partition\\ Skip hyperconnections} &\makecell[l]{Fault recovery ability}\\
\botrule
\end{tabular}
}
\end{table*}

\subsubsection{Total inference latency minimization}
Due to the large span of collaborative DNN inference system model and the need for mutual communication between different equipment, the total latency has always been a problem to be considered in collaborative inference. In cloud-edge-device collaborative DNN inference, in order to minimize the total inference latency, Ren \emph{et al.} \cite{ren2021fine} propose a computing partition mechanism of distributed DNN. Under the premise of satisfying the quality of service, the efficiency of inference is improved by coordinating the calculation of heterogeneous equipment. Xue \emph{et al.} \cite{xue2021eosdnn} propose a DNN inference accelerated offloading scheme in cloud-edge-device collaborative environment. It comprehensively considers large-scale model partition plan and migration plan, reduces inference latency and optimizes DNN real-time query performance. Chang-You Lin \emph{et al.} \cite{lin2019distributed} study the deployment of distributed DNN with limited completion time to solve the deployment problem considering both response time and inference throughput.Dey \emph{et al.} \cite{dey2019offloaded} realize a deep learning inference system which involved a robot vehicle based on Raspberry Pi 3 and hardware accelerator of Intel, reducing inference latency and improving tasks efficiency.

\subsubsection{Total cost minimization}
In the collaborative inference environment based on cloud, edge and terminal equipment, the system cost also needs to be considered. To reduce the system overhead caused by data transmission and hierarchical execution, Bing Lin \emph{et al.} \cite{lin2019cost} propose an adaptive particle swarm optimization algorithm. This method considers the characteristics of DNN partition and layers offloading to cloud, edge and terminal devices, which significantly reduces the system cost of DNN application offloading. Teerapittayanon \emph{et al.} \cite{teerapittayanon2017distributed} propose a distributed deep neural network (DDNN) at the level of distributed computing. DDNN maps the parts of a single DNN to the distributed computing hierarchy. While being able to adapt to the DNN inference on cloud, DDNN also allows fast and local inference using the shallow part of the neural network on edge and terminal devices. DDNN can not only achieve high accuracy, but also reduce the communication cost.

Liu and Peng \emph{et al.} \cite{liu2018edgeeye} propose \emph{EdgeEye}, an edge computing framework for real-time intelligent video analysis applications. \emph{EdgeEye} enables developers to transform the model trained by the popular deep learning framework into deployable components with minimal workload to optimize inference performance and efficiency. Tao \emph{et al.} \cite{tao2018esgd} consider a distributed deep learning framework based on cloud-edge-device, in which many edge devices cooperate in training models and use edge servers as parameter servers. However, the high network communication cost between edge devices and cloud is a bottleneck. The authors propose a new method called edge random gradient descent to reduce the communication cost of model parameters.

\subsubsection{Failure-resilient distributed DNN model}
In the distributed inference of neural network, the network is divided and distributed to multiple physical nodes. However, the failure of physical nodes will cause significant decline in inference performance of the system when the neural network is partitioned and distributed among them. Therefore, we need to consider the fault recovery ability of the system in cloud-edge-device collaborative inference framework, so as to obtain the failure-resilient distributed DNN model. If a device of the system is destroyed, the previous device can get an alternative path to transfer inference task and ensure the integrity of the system.

Yousefpour \emph{et al.} \cite{yousefpour2019guardians,yousefpour2020resilinet} introduce the concept of skipping hyperconnections in distributed DNN, which provides a certain fault recovery capability for inference in distributed DNN. The concept of skipping hyperconnections is similar to skipping connections in the residual network. It skips one or more physical nodes in the distributed neural network, forwards information to further physical nodes in the distributed structure, and provides an alternative path in case of physical node failure. Thus, the scheme can achieve flexibility for distributed collaborative DNN in cloud-edge-device network.

\subsection{Summary and analysis}
Compared with the previous two collaborative DNN inference paradigms, the research on cloud-edge-device collaborative DNN inference can also be considered as a supplement to cloud-device collaborative inference and edge-device collaborative inference, and it has serval advantages as follows:
\begin{enumerate}[1)]
\item \textbf{Higher resource utilization}. Through fine-grained division of the model, cloud-edge-device collaborative DNN inference can make full use of the resources of cloud center, edge nodes and terminal devices, which means higher resource utilization.
\item \textbf{Fault recovery capability}. Considering the fault recovery ability of large-scale IoT system, cloud-edge-device collaborative DNN inference establishes a failure-resilient distributed system model to enhance sensor fusion, system fault tolerance and data confidentiality.
\end{enumerate}

\section{Device-device collaborative DNN inference}\label{sec6}
In this section, we firstly introduce the motivation and architecture of device-device collaborative DNN inference paradigm. Then we summarize the performance optimization of the researches in recent years.Finally, we analyse the advantages and challenges of device-device collaborative DNN inference paradigm.

\subsection{Motivation and architecture}
\begin{figure}
\centerline{\includegraphics[width=\textwidth]{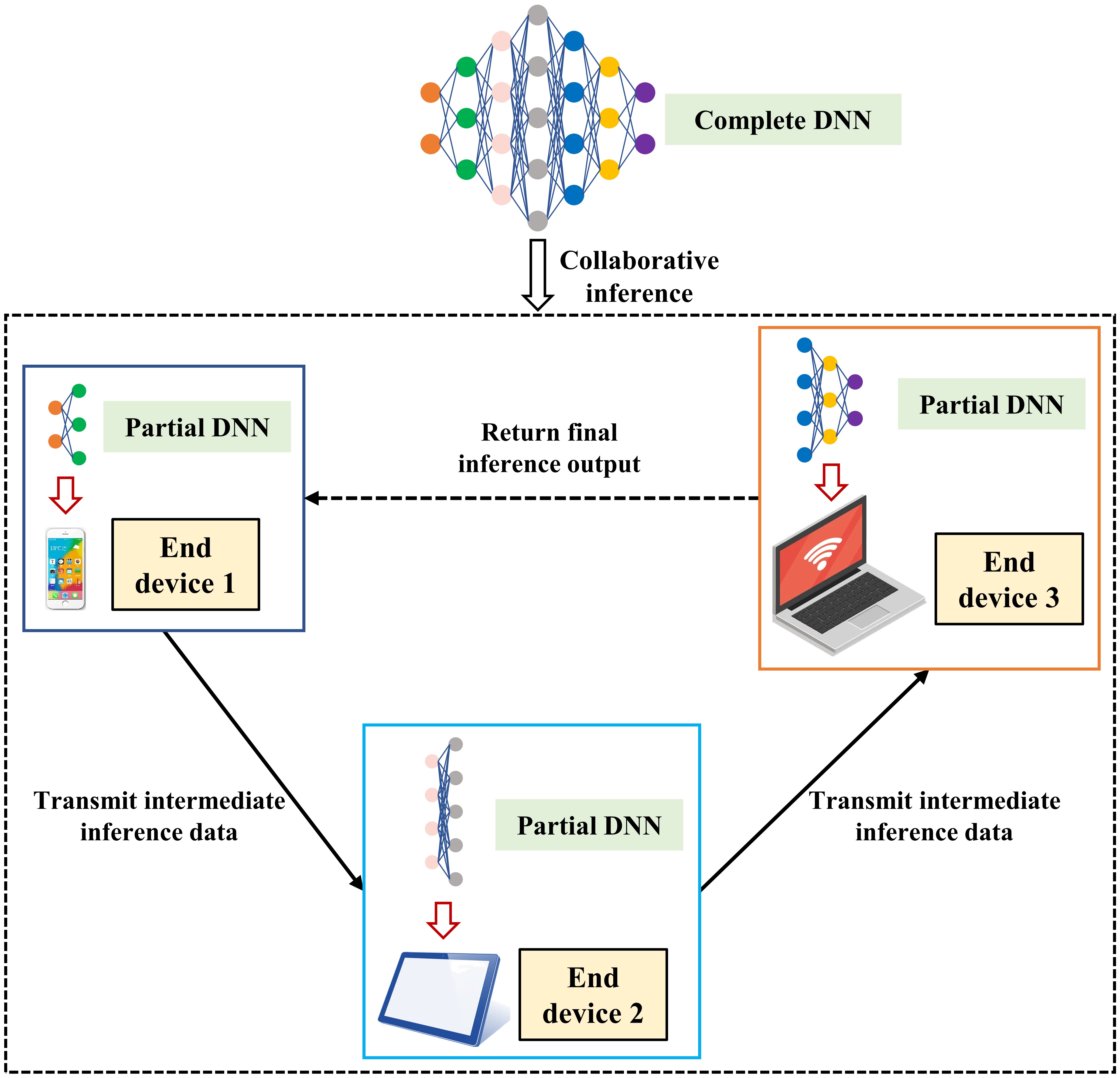}}
\caption{Illustration of device-device collaborative DNN inference.\label{fig7}}
\end{figure}

We note that in all previous work, an important scenario has not been fully explored, that is running DNN on a local distributed mobile computing system. Fig.~\ref{fig7} shows a framework of device-device collaborative DNN inference, with the development of mobile edge computing, more and more intelligent services and applications based on DNN are deployed on mobile terminal devices to meet the diversified and personalized needs of users. Compared with the client-server mode of a single mobile device supported by external infrastructure such as cloud, local device-device collaborative DNN inference computing system provides several important advantages, including more local computing resources, higher privacy, less network bandwidth dependence and so on. \emph{The paradigm can be used in high mobility scenarios or some remote and harsh environments, because terminal devices are low cost and they can be deployed everywhere. However, the resources of mobile devices are usually limited, including computing power and electric quantity, so the performance optimization objectives will focus on inference latency and energy consumption in device-device collaborative inference.}

\subsection{Performance optimization}
According to different performance optimization objectives, we divide device-device collaborative DNN inference into two categories, as shown in Table.~\ref{tab6}.

\begin{table*}[h]
\caption{Classification of device-device collaborative DNN inference}
\label{tab6}
\centering
\resizebox{\textwidth}{!}{
\begin{tabular}{ccccc}
\toprule
\textbf{Optimization Objective} & \textbf{Framework} & \textbf{Application} & \textbf{Main Technology} & \textbf{Effectiveness} \\
\midrule
\multirow{20}{1.5in}{Total inference latency minimization}
&DistInference \cite{dhuheir2021efficient} & Computer vision & \makecell[c]{Model partition} & \makecell[l]{Lower latency} \\
\cline{2-5}
&OULD \cite{jouhari2021distributed} & Computer vision &\makecell[c]{Model partition} &\makecell[l]{Lower latency\\Lower shared data} \\
\cline{2-5}
&\cite{disabato2021distributed} & IoT &\makecell[c]{Model partition} &\makecell[l]{Lower latency}\\
\cline{2-5}
&\cite{naveen2021low} &IoT &\makecell[c]{Model partition} &\makecell[l]{Communication size reduction: 14.4$\%$\\Inference latency reduction: 16$\%$}\\
\cline{2-5}
&\cite{du2020distributed} &IoT &\makecell[c]{-} &\makecell[l]{System acceleration: 3.85$\times$\\Memory footprint reduction: 70$\%$}\\
\cline{2-5}
&DistPrivacy \cite{baccour2020distprivacy} &IoT &\makecell[c]{Model partition} &\makecell[l]{Lower latency\\Privacy protection}\\
\cline{2-5}
&$\text{Edge}^{n}$AI \cite{hemmat2022text} & -&\makecell[c]{Model partiton\\Class-aware pruning} &\makecell[l]{Inference speedup: 17$\times$}\\
\cline{2-5}
&DeepSlicing \cite{zhang2021deepslicing} &Computer vision &\makecell[c]{Model partition\\Data partition} &\makecell[l]{Inference latency reduction: 5.79$\times$\\Memory footprint reduction: 14.72$\times$}\\
\cline{2-5}
&MoDNN \cite{mao2017modnn} & Computer vision &\makecell[c]{Data partition} &\makecell[l]{Computation acceleration: 2.17-4.28$\times$\\Data delivery time reduction: 30.02$\%$}\\
\cline{2-5}
&DeepThings \cite{zhao2018deepthings} & IoT&\makecell[c]{Fused tile partitioning\\Model partition\\Data partition} & \makecell[l]{Memory footprint reduction: 68$\%$\\Throughput improvement: 1.7-2.2$\times$\\Inference acceleration: 1.7-3.5$\times$}\\
\midrule
\multirow{6}{1.5in}{Total cost minimization}
&CoEdge \cite{zeng2020coedge} &IoT &\makecell[c]{Model partition\\Data partition} &\makecell[l]{Energy consumption reduction: 25.5-66.9$\%$}\\
\cline{2-5}
&\cite{hadidi2018distributed} &IoT &\makecell[c]{Model partition} &\makecell[l]{Lower memory footprint\\Lower energy consumption}\\
\cline{2-5}
&\cite{goel2022efficient}&Computer vision &\makecell[c]{Model partition} &\makecell[l]{Throughput improvement: 3.21$\times$\\Energy consumption reduction: 68$\%$\\Memory usage reduction: 58$\%$} \\
\botrule
\end{tabular}
}
\end{table*}

\subsubsection{Total inference latency minimization}
The equipment of device-device collaborative DNN inference is local and close to the data source. Although compared with the collaborative inference mode requiring remote server, its physical transmission distance is shorter and the communication latency will be smaller in theory. Due to the limited resources of local mobile devices, the latency of inference calculation will increase, and the increase of devices that need to participate in collaboration will also increase the communication latency. Therefore, it is necessary to study how to reduce the latency of device-device collaborative DNN inference.

UAV has attracted much attention because of its low cost, high mobility and excellent application ability in difficult and dangerous fields. However, the application of DNN on UAV has many challenges in dealing with deep networks and complex models. Dhuheir \emph{et al.} \cite{dhuheir2021efficient} propose a strategy to allocate inference requests to resource constrained UAV groups to classify the captured airborne images, so as to obtain the minimum decision latency. Jouhari \emph{et al.} \cite{jouhari2021distributed} propose a DNN distribution method in UAV to realize data classification in resource constrained equipment and avoid additional latency introduced by server based solution due to data communication from air to ground link.

Disabato \emph{et al.} \cite{disabato2021distributed} introduce a method designed to allocate CNN execution on Distributed IoT applications. The method is formalized as an optimization problem to minimize the end-to-end latency of collaborative inference under given memory and load constraints. Naveen \emph{et al.} \cite{naveen2021low}  propose a device-device edge computing framework, which uses the weight pruning method to promote optimization, deploys the model to low performance intelligent devices designed for real-time applications, and has made significant improvements in communication scale and inference latency.

For the application of IoT, Du and Jiangsu \emph{et al.} \cite{du2020distributed} propose a distributed CNN inference system based on loosely coupled CNN structure, synchronous partition and asynchronous communication. The system reduces the memory occupation of each device and improves the efficiency of collaborative inference. Baccour \emph{et al.} \cite{baccour2020distprivacy} explore the allocation of DNN on the IoT devices in the monitoring system to minimize the latency of classification decision, who also introduce a model called \emph{distprivacy} to improve the system privacy. Hemmat \emph{et al.} \cite{hemmat2022text} propose a framework to decompose the complex DNN into multiple available local edge devices, which minimizes the communication overhead and overall inference latency.

Zhang \emph{et al.} \cite{zhang2021deepslicing} introduce \emph{Deepslicing}, a cooperative adaptive inference system, which is suitable for all kinds of CNNs and supports customized flexible fine-grained scheduling. Through the scheduler to the model and data, the balance between calculation and synchronization is realized, and the inference latency and memory occupation are reduced. Mao \emph{et al.} \cite{mao2017modnn} propose a local distributed mobile computing system named \emph{MoDNN} for DNN applications. \emph{MoDNN} can distribute the trained DNN model to multiple mobile devices, and accelerate DNN computing by reducing device level computing and memory usage. Zhao \emph{et al.} \cite{zhao2018deepthings} propose a lightweight framework named \emph{Deepthings} for adaptively distributed execution of CNN based inference applications on the edge cluster of the IoT with strictly limited resources. \emph{Deepthings} not only realizes the parallelism of independently distributed processing tasks, but also minimizes the memory occupation and reduces the overall inference latency.

\subsubsection{Total cost minimization}
Local mobile devices not only have limited computing and memory resources, but also have limited battery capacity due to volume constraints. The energy consumption of collaborative inference and communication will affect their activity range and task duration to a certain extent. Therefore, we need to consider how to minimize the total cost of device-device collaborative inference in application scenarios to improve task performance. Goel \emph{et al.} \cite{goel2022efficient} verify that the hierarchical DNN architecture is very suitable for parallel processing on multiple edge devices, and created a parallel inference system for computer vision problems of hierarchical DNN. The method balances the load between cooperative devices and reduces the communication cost, so as to process multiple video frames at the same time with higher throughput.

Zeng \emph{et al.} \cite{zeng2020coedge} propose \emph{CoEdge}, a distributed DNN computing system that coordinates heterogeneous edge devices for collaborative DNN inference. \emph{CoEdge} employs the computing and communication resources available at the edge to dynamically divide the DNN inference workload according to the computing power of the equipment and network conditions, which greatly reduces the energy consumption of model inference. Hadidi \emph{et al.} \cite{hadidi2018distributed} develop an analysis technology that can effectively distribute DNN model based inference applications on distributed robot systems, taking memory usage, communication overhead and real-time data processing performance into account.

\subsection{Summary and analysis}
Compared with the previous three collaborative inference methods, the device-device collaborative DNN inference has three prominent advantages:
\begin{enumerate}[1)]
\item \textbf{System independence}. It reduces the dependency between mobile devices and servers and reduces the operation cost of devices.
\item \textbf{Lower inference latency}. Device-device collaborative mode reduces the latency of inference decision which contains the local mobile devices communication with each other, receiving requests and making final decisions, so as to avoid the overhead of remote transmission.
\item \textbf{System adaptability}. When the remote server cannot be connected in a harsh environment or some devices are destroyed, the mobile device cluster can still perform inference tasks under the device-device collaborative mode.
\end{enumerate}

However, according to the existing research, device-device collaborative DNN inference still faces many challenges:
\begin{enumerate}[1)]
\item \textbf{Privacy security}. Due to the limited computing resources of local distributed devices, a single machine cannot execute a large DNN model. Multi-device collaborative inference will increase the communication overhead and the probability of data loss, and there is a risk of privacy disclosure in the frequent transmission of data.
\item \textbf{Synchronization of heterogeneous devices}. Due to the heterogeneity of local mobile devices and different computing power of different devices, it may lead to the idle waiting of high computing power devices during data synchronization, which reduces the inference efficiency and increases the end-to-end latency.
\item \textbf{Dynamic task allocation based on device status}. Due to the limitation of battery capacity, for collaborative inference tasks, fine-grained device scheduling strategy and task allocation strategy are needed to minimize the overall energy consumption. Only the matching between the remaining power of each device and the amount of tasks can ensure the complete operation of the whole inference system.
\end{enumerate}

\section{Future development trend of collaborative DNN inference for EI}\label{sec7}
\begin{figure}[h]
\centerline{\includegraphics[width=3.5in]{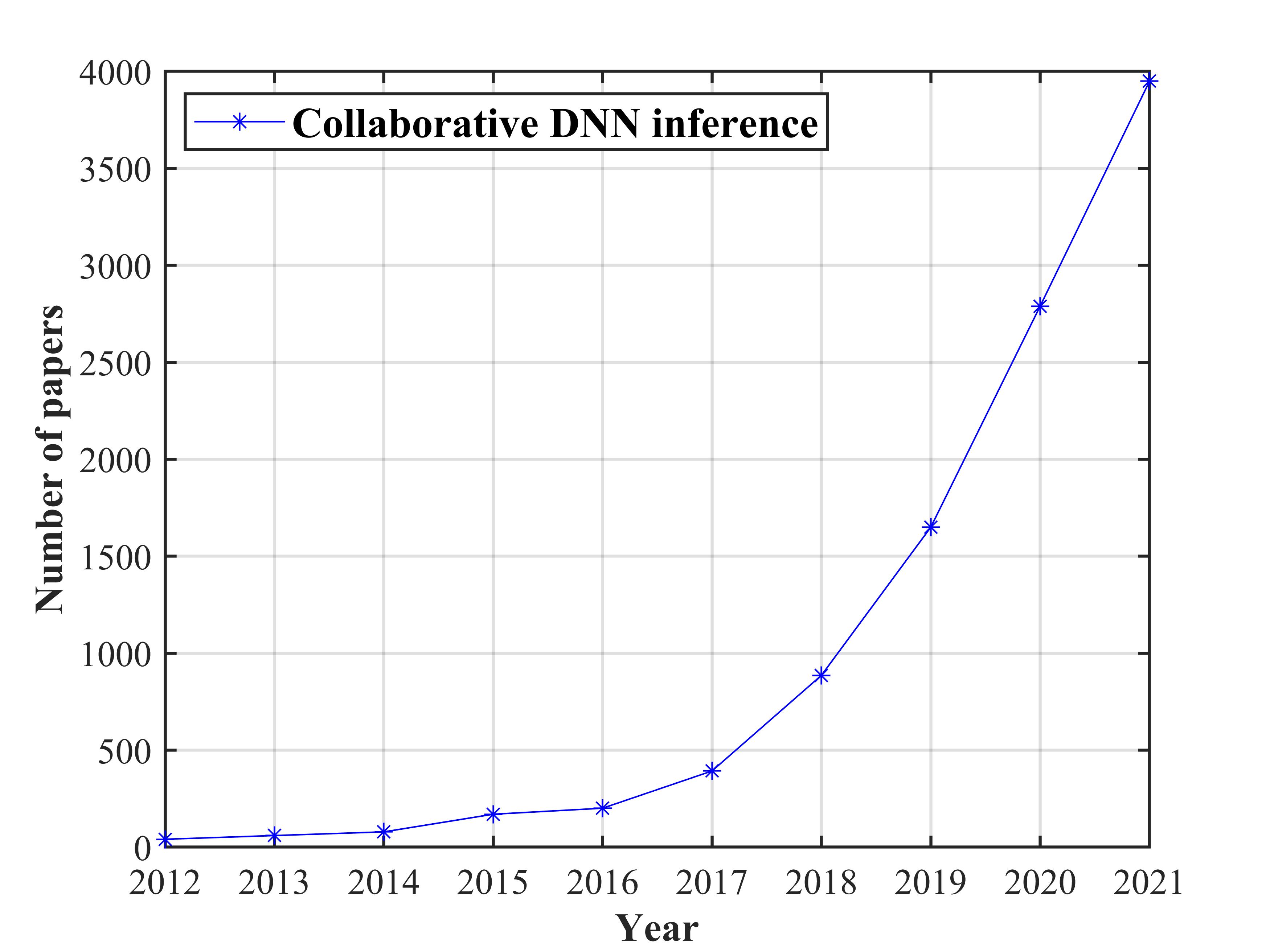}}
\caption{Number of papers related to by ``collaborative DNN inference" on Google Scholar.\label{fig8}}
\end{figure}

Based on the above comprehensive discussion of existing works, we will describe several open challenges and future research directions of collaborative DNN inference for EI in this section. In Fig.~\ref{fig8}, we count the number of papers related to ``collaborative DNN inference'' in recent years, which shows the rapid development and broad exploration space of this research field. With the emergence of AI driven computing intensive mobile and IoT applications, edge-oriented collaborative DNN inference can become a common model, but the current collaborative DNN inference paradigm still needs to be further explored to obtain higher performance improvement and wider scene applications.

\subsection{Trade-off design of performance index of collaborative DNN model}
For EI oriented collaborative DNN inference applications with specific tasks, there are generally many DNN models that can complete tasks as candidates. However, how to select an appropriate DNN model for collaborative DNN inference application is a research difficulty, because standard performance indicators such as top-k accuracy or average accuracy can not fully reflect the performance of collaborative DNN model when it runs cooperatively on edge devices. In the model deployment stage, in addition to accuracy, inference efficiency and resource consumption are also key indicators \cite{huang2017speed}. Therefore, we need to explore the trade-offs between these performance indicators and determine the factors affecting them, so as to obtain the optimal collaborative model selection and deployment mode of the current task.

\subsection{Dynamic computing network technology}
For EI oriented collaborative DNN inference, computing intensive applications usually run in distributed edge computing environment. Therefore, an advanced network solution with computing awareness is very necessary, so that the calculation results and data can be efficiently shared among different edge nodes. In 5G networks, the flexible control of network resources can support the on-demand interconnection between different edge nodes in computing intensive AI applications. For the future 6G networks, due to the addition of satellite communication, IoT can realize the global coverage of mobile communication, so the connection between collaborative DNN inference devices will be closer. On the other hand, the autonomous network mechanism that allows the dynamic configuration of new edge nodes and devices is important for efficient collaborative inference in dynamic heterogeneous networks. Collaborative inference system with dynamic autonomous network mechanism can also deal with the sudden failure of nodes, which is the research direction of failure-resilient collaborative model in the future.

\subsection{Intelligent task allocation and resource management}
Due to the distributed character of collaborative DNN inference, the edge devices and nodes participating in the intelligent task are scattered in different regions, and different edge nodes may run different DNN models. Therefore, it is important to schedule the equipment reasonably, make full use of the scattered resources between the edge nodes and the equipment, divide the complex model into multiple subtasks, and offload these tasks efficiently between the edge nodes and the equipment, so as to perform inference cooperatively.

The environment of collaborative DNN inference application scenarios is highly dynamic, and it is difficult to accurately predict future events. Therefore, it needs excellent online edge resource coordination and provisioning capabilities to adapt to large-scale tasks. The real-time joint optimization of heterogeneous computing, communication and cache resource allocation, as well as the customized system parameter configuration for different task requirements are also important. In order to solve the complexity of algorithm design, an emerging research direction is the efficient resource allocation strategy of data-driven adaptive learning.

\subsection{Security and privacy issues}
Because distributed collaborative inference needs to ensure that the services of inference tasks provided by different nodes are credible, the design of distributed security mechanism is significant to ensure the authentication of subscriber, the access control of collaborative inference tasks, the model and data security of devices, and the mutual authentication between different devices \cite{du2018big}. In addition, considering the coexistence of trusted edge nodes and untrusted edge nodes, it is also important to study new secure routing schemes and trusted network topology for collaborative DNN inference.

Terminal devices will generate a large amount of data at the edge of the network, which may involve privacy issues \cite{li2018klra}, because they may contain user location information or activity records. According to the requirements of privacy protection, directly sharing the raw data set among multiple nodes may have a high risk of privacy disclosure. Therefore, device-device collaborative DNN inference is a feasible paradigm. The raw data set is stored on the generation equipment, and only the extracted intermediate feature information and model parameters are transmitted between the local inference equipment groups. Meanwhile, there are also studies using differential privacy, homomorphic encryption and secure multi-party computing tools to design a privacy protected model parameter sharing scheme, which can further enhance the data privacy in collaborative inference for EI. Nowadays, there are also researches on collaborative inference to enhance the security and privacy of devices and data by combining blockchain technology, which is also one of the research directions that can be considered in the future collaborative DNN inference on privacy issues.

\section{Conclusion}\label{sec8}
This paper combs and classifies the research progress of edge collaborative DNN inference in recent years. Specifically, we firstly review the background and motivation of collaborative DNN inference at the edge of the network. Then, we outline the overall architecture, performance optimization and key technologies of four different types of collaborative DNN inference models. Finally, we discuss the open challenges and future research directions of collaborative DNN inference for EI. It is hoped that this survey can attract more and more attention, stimulate extensive discussion, and provide ideas for further research on collaborative DNN inference.

\bibliography{ref}

\end{document}